\documentclass[%
 reprint,
 amsmath,amssymb,
 aps,twocolumn,
 superscriptaddress
]{revtex4-1}

\usepackage{graphicx}
\usepackage{dcolumn}
\usepackage{bm}

\usepackage{color}
\begin{document}

\preprint{APS/123-QED}


\title{Enhanced relativistic-electron beam collimation using two consecutive laser pulses}


\author{S.$\,$Malko}
\affiliation{Centro de Laseres Pulsados (CLPU), Parque Cientifico, E-37185 Villamayor, Salamanca, Spain}
\affiliation{University of Salamanca, Salamanca, Spain}

\author{X.$\,$Vaisseau}
\affiliation{Centro de Laseres Pulsados (CLPU), Parque Cientifico, E-37185 Villamayor, Salamanca, Spain}

\author{F.$\,$Perez}
\affiliation{Laboratoire pour l'Utilisation des Lasers Intenses, Ecole Polytechnique, CNRS, CEA, UMR 7605, F-91128, Palaiseau, France}

\author{D.$\,$Batani}
\affiliation{Univ. Bordeaux, CNRS, CEA, CELIA (Centre Lasers Intenses et Applications), UMR 5107, F-33405 Talence, France}
\affiliation{National Research Nuclear University (MEPHI), Moscow, Russia}

\author{A.$\,$Curcio}
\affiliation{Laboratori Nazionali di Frascati (INFN), Frascati 00044, Italy}

\author{M.$\,$Ehret}
\affiliation{Univ. Bordeaux, CNRS, CEA, CELIA (Centre Lasers Intenses et Applications), UMR 5107, F-33405 Talence, France}

\author{J.J.$\,$Honrubia}
\affiliation{ETSI Aeron\'{a}uticos, Universidad Polit\'{e}cnica de Madrid, Madrid, Spain}

\author{K.$\,$Jakubowska}
\affiliation{IPPLM, Warsaw, Poland}

\author{A.$\,$Morace}
\affiliation{Institute of Laser Engineering, Osaka University, 2-6 Yamadaoka, Suita, Osaka 565-0871, Japan}

\author{J.J.$\,$Santos}
\affiliation{Univ. Bordeaux, CNRS, CEA, CELIA (Centre Lasers Intenses et Applications), UMR 5107, F-33405 Talence, France}

\author{L.$\,$Volpe}
\email{lvolpe@clpu.es}
\affiliation{Centro de Laseres Pulsados (CLPU), Parque Cientifico, E-37185 Villamayor, Salamanca, Spain}
\affiliation{University of Salamanca, Salamanca, Spain}

\date{\today}


\begin{abstract}
The double laser pulse approach to relativistic electron beam (REB) collimation has been investigated at the LULI-ELFIE facility. In this scheme, the magnetic field generated by the first laser-driven REB is used to guide a second delayed REB. We show how electron beam collimation  can be controlled by properly adjusting laser parameters. By changing  the ratio of focus size and the delay time between the two pulses we found a maximum of electron beam collimation clearly dependent on the focal spot size ratio of the two laser pulses and related to the magnetic field dynamics. Cu-K$_{\alpha}$ and CTR imaging diagnostics were implemented to evaluate the collimation effects on the respectively low energy ($\leq$ 100 keV) and high energy ($\geq$ MeV) components of the REB.


\begin{description}
\item[PACS numbers]
\end{description}
\end{abstract}

\pacs{Valid PACS appear here}
\maketitle

The study of the transport of relativistic laser-driven electrons is a subject of interest for many applications including proton-ion acceleration \cite{clark2000,park2006}, fast ignition approach to inertial confinement fusion (ICF) \cite{tabak1994,eidmann2000} , astrophysics applications \cite{perez2010} as well as high brilliance and compact laser-based x-ray sources \cite{morace2012,antonelli2014}. In the fast ignition approach to ICF an ultra-intense laser is used to produce relativistic electrons which deposit their energy to ignite a pre-compressed fuel pellet. This scheme requires specific conditions for the laser-electrons conversion efficiency ( $\geq 40\%$) and  for the mean energy ($ \sim 1 - 2$  MeV) of the electrons delivered into the 20 $\mu$m radius core \cite{atzeni2009}. Reducing electron beam divergence \cite{green2008} as well as optimising electron beam transport in plasmas  is crucial in order to satisfy these drastic conditions. Previous investigations have shown that the dynamics of electron beams propagation in plasmas is mainly affected by:  i) resistivity effects \cite{volpe2013a, volpe2013b,vauzour2012,santos2007,vaisseau2015,santos2017,pisani2000,batani2000,batani2002} on the electron stopping power, which become important at relativistic intensities ( $I_L \geq 10^{18}\,$W.cm$^{-2}$) and reduce the final penetration length of the electron beam; ii) collisionless Weibel instabilities which starts to grow and become very important for laser intensities $I_L>10^{19}\,$W.cm$^{-2}$, at the level of the plasma skin depth, generating micro magnetic fields that strongly contribute to increase the initial electron divergence \cite{debayle2010}.
Different strategies to control REB propagation in solid matter have been proposed. They rely on the use of $\sim$kT magnetic fields, which can be externally generated by coils \cite{santos2015, bailly2018} or self generated \cite{bell2003}, either by artificial resistivity gradients \cite{kar2009, schmitz2012, debayle2013,santos2013,Perez2011} or by exploiting the intrinsically high resistivity of a material \cite{vaisseau2017}. An alternative scheme by using self-generated magnetic fields was proposed by A. Robinson \textit{et al.} \cite{robinson2008,robinson2014}. In this scheme two collinear laser pulses ($1$ and $2$) with a given intensity ratio ${I_2}/{I_1}$ $\sim$ 10 separated by a delay ($\Delta t = t_2-t_1$) are used to generate fast electron beams. The electron beam produced by the first, less intense laser pulse generates a resistive azimuthal magnetic field (seed magnetic field) to guide the main electron population generated by the second beam. This scheme was experimentally investigated by Scott et al. \cite{scott2012} who have shown the existence of an optimum delay between the laser pulses of the order of the laser pulse duration  ($\Delta t \sim \tau$), at which a maximum electron beam collimation is reached. The existence of this optimum can be explained by considering the growth rate and then the dynamics of the spatial diffusion of the seed magnetic field, in connection with the arrival time of the main electron beam. Further analysis of results have been underlined by numerical simulations and theoretical predictions \cite{volpe2014}. Although a promising collimation effect was observed in \cite{scott2012}, the study was focused on the only influence of the delay time between the laser pulses while other relevant parameters, namely laser intensity and laser focal spot sizes ratio, were kept constant. 

Indeed, the dynamics of the seed magnetic field mainly depends on electron current density and the consequent evolution of resistivity in the target. Since these can be easily controlled by modifying the laser focal spot size, it is important to study the relation between the radius ${R_1}$ of the azimuthal magnetic field created by the first beam and the radius of the second electron beam. With respect to this Robinson \textit{et al.} \cite{robinson2008} has suggested (by assuming the Larmor radius of the second beam smaller than the radial extension of the seed magnetic field)  that the best condition for collimation can be written in terms of the laser focal spots ratio ${\varphi_1} >{\varphi_2 }$ . 

\begin{figure}[!htbp]
\begin{center}
\includegraphics[width=0.9\linewidth]{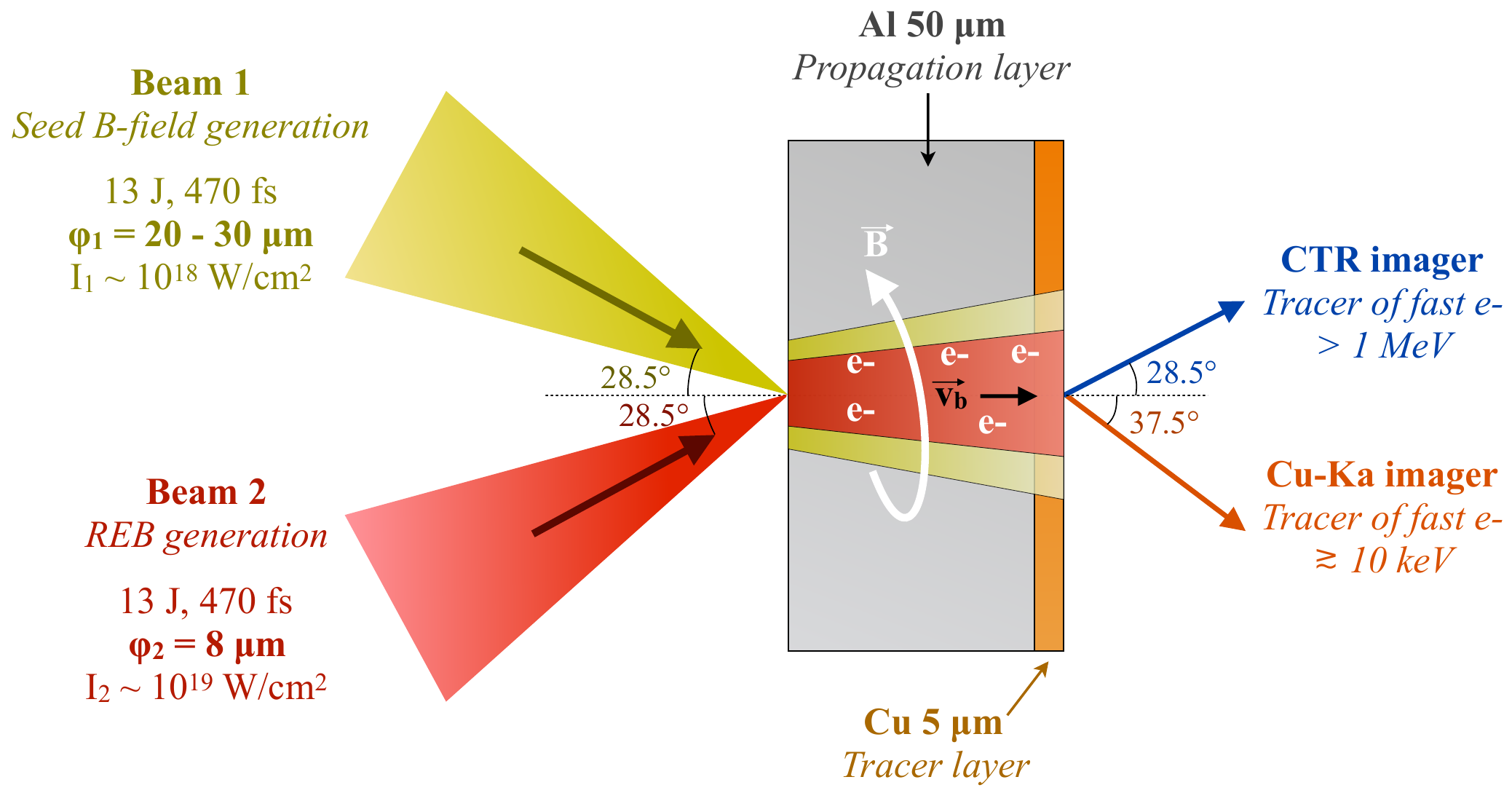}
\end{center}
\caption{\small (color online) Experimental setup. }
\label{fig:Setup}
\end{figure}

\begin{figure}[!htbp]
\begin{center}
\includegraphics[width=0.9\linewidth]{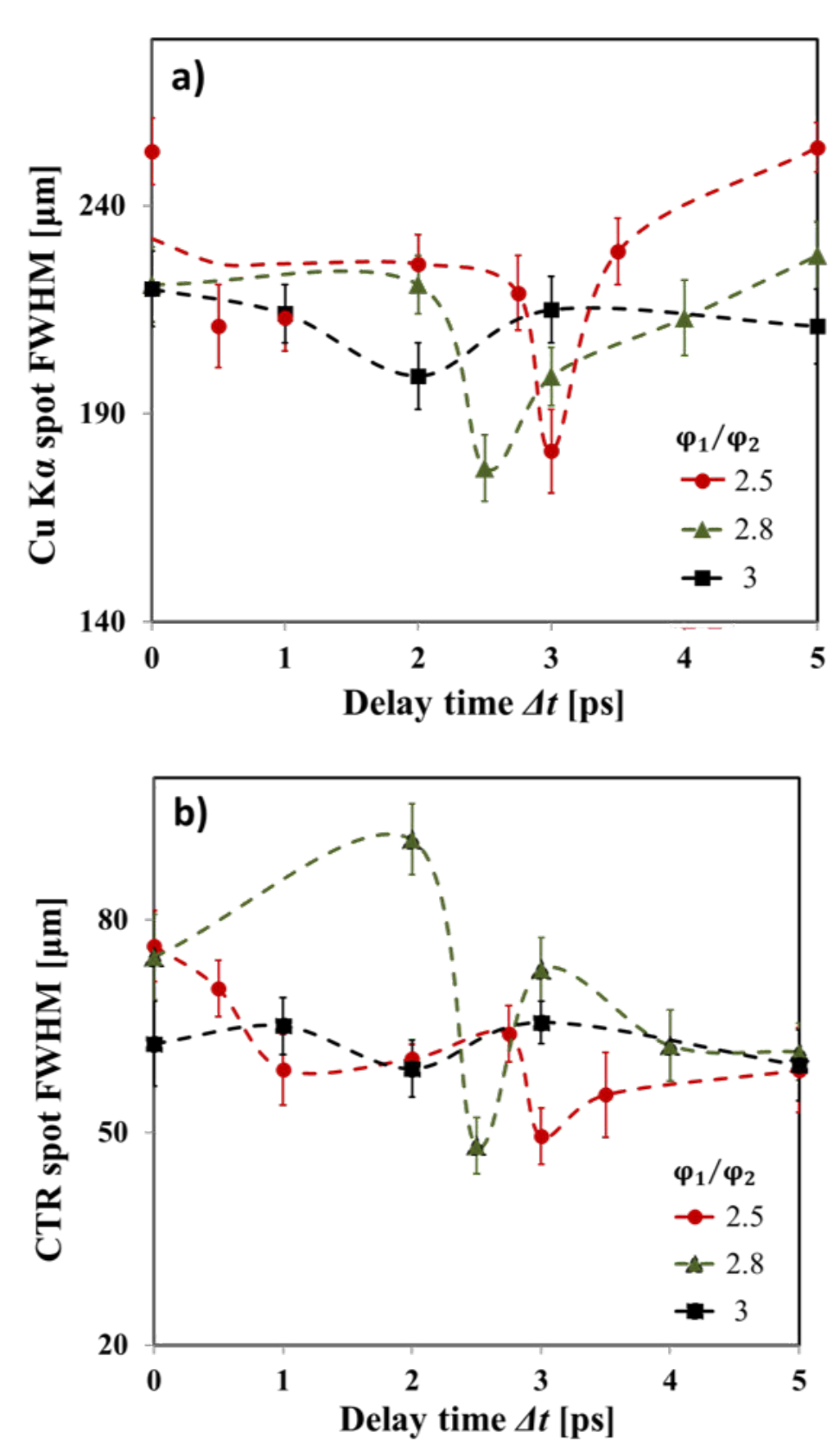}
\end{center}
\caption{\small (color online)Evolution of the diameter of the emission area on target rear side of {\bf (a)} Cu-K$_{\alpha}$ fluorescence and {\bf (b)} CTR, as a function of the delay between the two laser pulses for different focal spot ratios: ${\varphi_{1}}/{\varphi_{2}}=2.5$ (red circles), ${\varphi_{1}}/{\varphi_{2}}=2.8$ (green triangles), ${\varphi_{1}}/{\varphi_{2}}=3$ (black squares). The error bars are estimated by the standard deviation from multiple shots taken in the same laser conditions. The dashed curves are guides for the eyes.}
\label{fig:KaGOI}
\end{figure}

\begin{figure}[!htbp]
\centering
\includegraphics[width=0.9\linewidth]{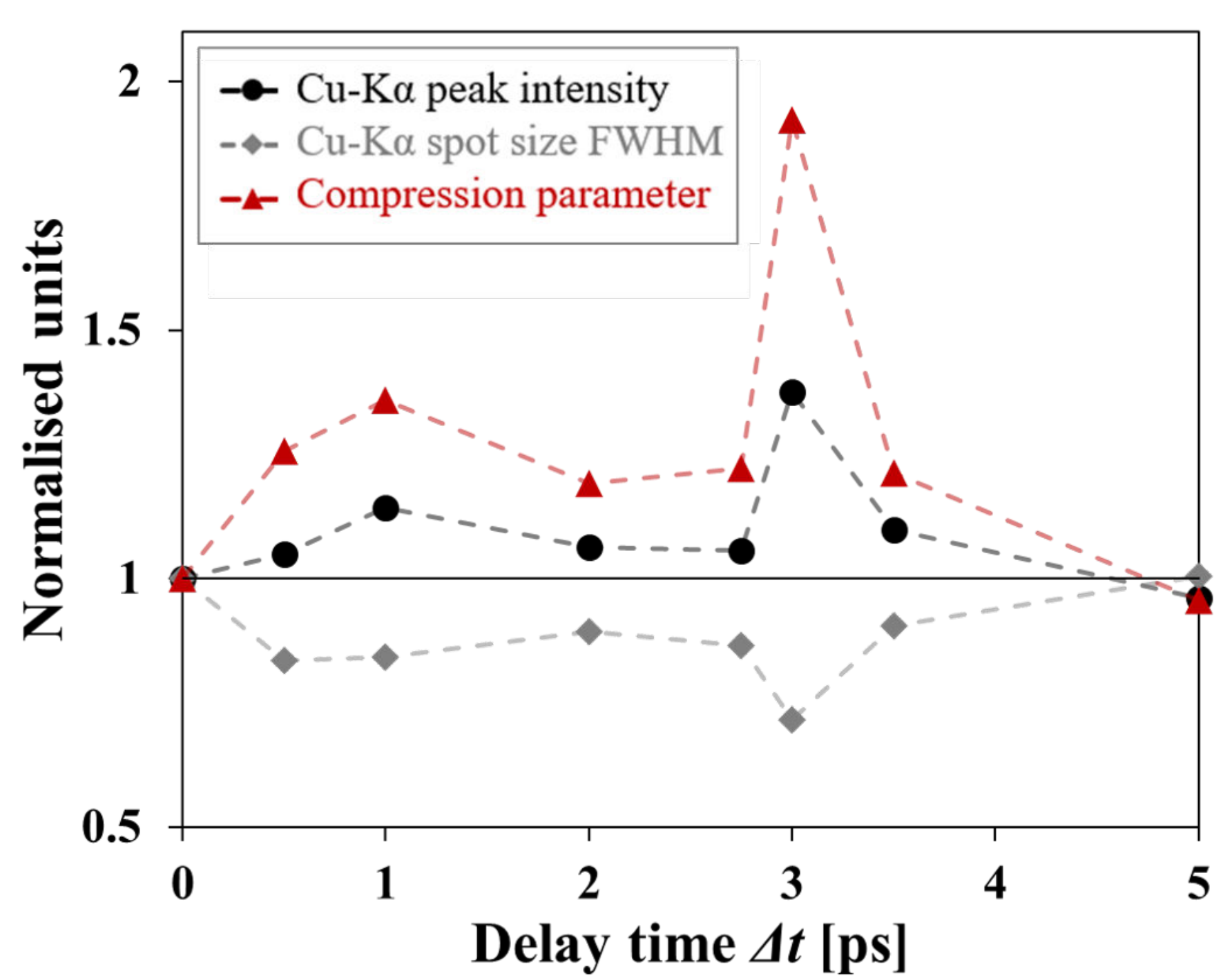}
\caption{\small (color online) Evolution of the Cu-K$_{\alpha}$ peak intensity (black circles), Cu-K$_{\alpha}$ emission spot size (grey circles) and compression factor C (red triangles), normalized to the values at $\Delta t = 0\,$ ps for the run with the focal spot ratio ${\varphi_{1}}/{\varphi_{2}}=2.5$. 
\label{fig:Compression}}
\end{figure}

In this context we report the results of a novel experimental scheme of fast electron collimation that uses two independent focusing parabolic mirrors, allowing to vary the ratio ${\varphi_{1}}/{\varphi_{2}}$  between the two laser focal spots, therefore controlling the ratio $R_1/R_2$. In addition to the Cu-K$_{\alpha}$ emission diagnostic used in \cite{scott2012} and mainly sensitive to the more numerous electrons in the 10 - 100 keV range, we implemented measurement of coherent transition radiation (CTR) to evaluate the collimation effect on higher energy electrons ($\gtrsim$ 1 MeV) \cite{baton2003}. The high number of shots performed in the campaign allowed obtaining better statistical characterisation of the collimation efficiency.  
%
%
%
%
%
%
\\
The experiment was performed at the ELFIE facility (Ecole Polytechnique, France). We used a dual beam configuration, with two $\lambda = 1.06\,\mu$m, $\tau$ = 470 fs full-width-at-half-maximum (FWHM) pulses, containing 13 $\pm$ 2 J of energy each and focused symmetrically at $\pm\,\ 28.5^0$ incidences with respect to the target normal [Fig. \ref{fig:Setup}] . The use of two different off-axis parabolic mirrors, one for each beam, allowed to vary the focal spot size of the first pulse, generating the seed magnetic field, from $\varphi_1 = 20\,\mu$m to 30 $\mu$m FWHM while keeping constant the focal spot of the second pulse ($\varphi_2 = 8\,\mu$m FWHM). These yielded intensities of $\sim10^{18}\,$W.cm$^{-2}$ and $1\times10^{19}\,$W.cm$^{-2}$, respectively. The $3\times3\,$mm$^2$ planar targets were composed of Al[50$\,\mu$m] - Cu[5$\,\mu$m], with the Al layer facing the two laser pulses. The two pulses, originating from the same oscillator, were temporally synchronized using interferometry techniques. The delay $\Delta t$ between the laser pulses was varied between $0\,$ps and $5\,$ps, with a precision of $100\,$fs.
Cu-K$_{\alpha}$ x-ray emission ($ \hbar\omega \approx 8 $ keV) produced by electrons passing through the copper tracer was imaged onto a FUJI image plate using a spherically bent quartz 22-43 crystal \cite{morace2012,antonelli2014}, with a radius of curvature of 25 mm, looking at $37.5^0$ with respect to the target normal. Coherent transition radiation at twice the laser frequency \cite{santos2002} produced by relativistic electrons of energies $\gtrsim1$ MeV was recorded using a Gated Optical Imager (GOI) looking at the target rear surface at $28.5^0$ with respect to the target normal and with an acquisition time of 200 ps, limiting the contribution to the signal of delayed Planckian thermal radiation.
\\
Figure \ref{fig:KaGOI}(a)  presents the evolution of the Cu-K$_{\alpha}$ spot size as a function of delay time between the two laser pulses for different ratios ${\varphi_{1}}/{\varphi_{2}}$. An optimum delay, corresponding to a maximum collimation of the fast electron beams, was measured for each focal spot ratio. Both Cu-K$_{\alpha}$ and CTR diagnostics  confirm  that  the collimation of the main electron beam occurs at delays $\Delta t=3 ,2.5,2 $ ps respectively  for the run with focal spot ratios   ${\varphi_{1}}/{\varphi_{2}}=2.5 ,2.8, 3$ [Fig.\ref{fig:KaGOI} (b)]. The higher the ratio ${\varphi_{1}}/{\varphi_{2}}$, the shorter is the delay at which an optimum collimation occurred. 
\\
\\
\\
\begin{figure*}[!htbp]
\begin{center}
\includegraphics[width=0.8\linewidth]{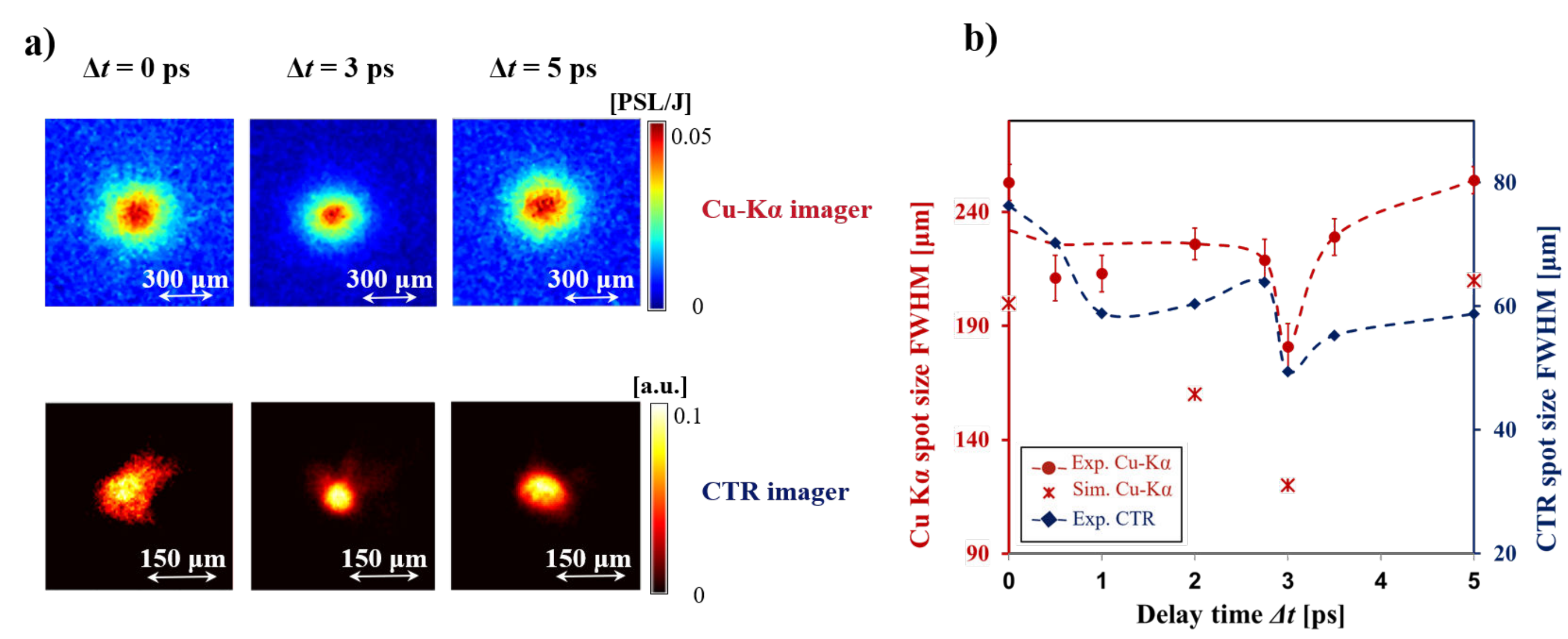}
\end{center}
\caption{\small (color online) Data obtained for a focal spot ratio $\varphi_{1}/\varphi_{2}$ = 2.5. {\bf (a)} Set of typical Cu-K$_{\alpha}$ (top) and CTR (bottom) images obtained at different delays $\Delta t = 0\,$ ps (left), 3 ps (middle) and 5 ps (right). {\bf (b)} Comparison of Cu-K$_{\alpha}$ (red circles) and of CTR (blue circles) emission spot sizes. The red crosses show the results of the simulated Cu-K$_{\alpha}$ emission, reproducing the delay at which optimal collimation occurs.}
\label{fig:Scan2_image}
\end{figure*}
\\
Examples of experimental images are shown in figure \ref{fig:Scan2_image}(a) for ${\varphi_{1}}/{\varphi_{2}}=2.5$, while figure \ref{fig:Scan2_image}(b) reports the size as measured by the two diagnostics on the same graph [same data as figure \ref{fig:Scan2_image}(a)]. Compared to the K$_{\alpha}$ signals the absolute smaller size of CTR signal confirms that this emission is due to the high energy component of electron beam, which has a smaller angular spread.
\noindent  We now introduce the electron beam compression parameter C defined as the ratio between the Cu-K$_{\alpha}$ spot FWHM and the Cu-K$_{\alpha}$ peak intensity \cite{volpe2014}. A compression of the beam is indeed achieved when a reduction of the electron beam size is accompanied by an increase of the peak intensity of the signal: a smaller value of C corresponds to a more collimated electron beam. 
Fig. \ref{fig:Compression}(a) shows the evolution, as a function of the delay between the laser pulses $\Delta t$, of both the Cu-K$_{\alpha}$ peak intensity (black circles) and spot size (grey diamonds) [both normalised to their value at $\Delta t$ = 0] and of the compression parameter C (red triangles) for the case $\varphi_{1}/\varphi_{2}$ = 2.5. The maximum compression corresponds to the maximum value of Cu-K$_{\alpha}$ emission at the delay time of 3 ps where the electron beam area is decreased by a factor of 0.5 and the Cu-K$_{\alpha}$ intensity is increased by a factor of 1.37. This suggests that more than 70\% of electrons are collimated in the process. The CTR signal shows also a reduction of the beam size by a fairly comparable factor $\sim$ 0.6 although there is not a clear increase of the detected signal yield. This seems to suggest a lower effect of the magnetic field on the high energy electron beam component which results both because of  the larger difference between the radial extent of the magnetic field and the spatial size of the high energy component in the beam, and the smaller deviation of higher energy electrons. 
\noindent Two main effects were observed when varying the laser focal spot ratio of the laser pulses: a variation of the maximum compression coefficient and a shift of the optimum delay time. Such tendencies are explained by the dynamics of the self-generated magnetic fields governed by the diffusion equation whcih combine the generalized Ohm$'$s law \cite{braginski1965,nicolai2011} with Maxwell-Faraday$'$s law:
\\
\begin{equation}\label{Faraday}
\dfrac{\partial\boldsymbol{B}}{\partial t} = \eta \boldsymbol{\nabla} \times \boldsymbol{j_b} + \boldsymbol{\nabla} \left(\eta \right) \times \boldsymbol{j_b} + \dfrac{\eta}{\mu_0}\boldsymbol{\nabla}^2\boldsymbol{B}-  \dfrac{1}{\mu_0} \boldsymbol{\nabla} \left(\eta \right) \times \boldsymbol{B} 
\end{equation}
\\
Here $\eta$, $\boldsymbol{B}$ are the plasma resistivity and magnetic field respectively, $\boldsymbol{j_{b}}$ is the fast electron current density. The terms in the right-hand-side of Eq. \ref{Faraday} are responsible for the magnetic field generation and evolution. The maximum amplitude $B_{max}$, the rise and diffusion times are mainly dependent on the laser pulse duration, intensity  and focal spot size via the target resistivity evolution and the fast electron beam current density. Applying this equation to our case, we can explain the magnetic field dynamics and its influence on the observed electron beam collimation. An increase of the focal spot size of the first laser pulse causes (see Fig. \ref{fig:KaGOI}):

 1) A reduction of the optimum delay time
 
 2) An increase of the time window for second electron beam injection 
 
 3) A mitigation of the REB collimation.  
 
\noindent The later effect (3), estimated by the compression ratio, is caused by the natural reduction of the maximum amplitude of magnetic field  $B_{max}$  because a larger spot implies a reduced laser intensity on target: the $\Delta t$ scan with ${\varphi_{1}}/{\varphi_{2}}=3$ is the less efficient.

\noindent The reduction of the optimum delay time between two laser pulses when the focal spot ratio increased (1) is due to the change in target resistivity following the evolution of target  temperature. With the increase of laser focal spot $\varphi_1$ , the injected energy density reduces, therefore the target electron temperature $T_e$ decreases and the resistivity $\eta$ gets larger implying a decrease of the B-field rise time $\xi \sim \frac {1} {\eta}$.  As a consequence the magnetic field reaches $B_{max}$ faster, when the REB collimation is observed. 
\\
\noindent As for the optimum time window (2) for the injection of the second electron beam, this appears because the collimation of the REB is caused by a resistive magnetic field presenting a sufficiently long lifetime, the later being directly related to the magnetic field diffusion time, scaling as $\tau_{\textrm{diff}} \propto \frac {R^2} {\eta}$ [Eq. \ref{Faraday}]. As a consequence, the bigger the radial size of the first electron beam,  the longer the seed magnetic field lasts, extending the optimum time window for the injection of the main electron population. 
The existence of the optimum focal spot ratio, when the compression reaches its maximum, is a trade-off between the maximum amplitude of the magnetic field $B_{max}$ and its diffusion time. The laser focal spot ratio should lay between $2 - 2.8$,  the most evident collimation effect having been observed for a  focal spot ratio of 2.5.
\\
\\
\\
\begin{figure}[!htbp]
\centering
\includegraphics[width=0.9\linewidth]{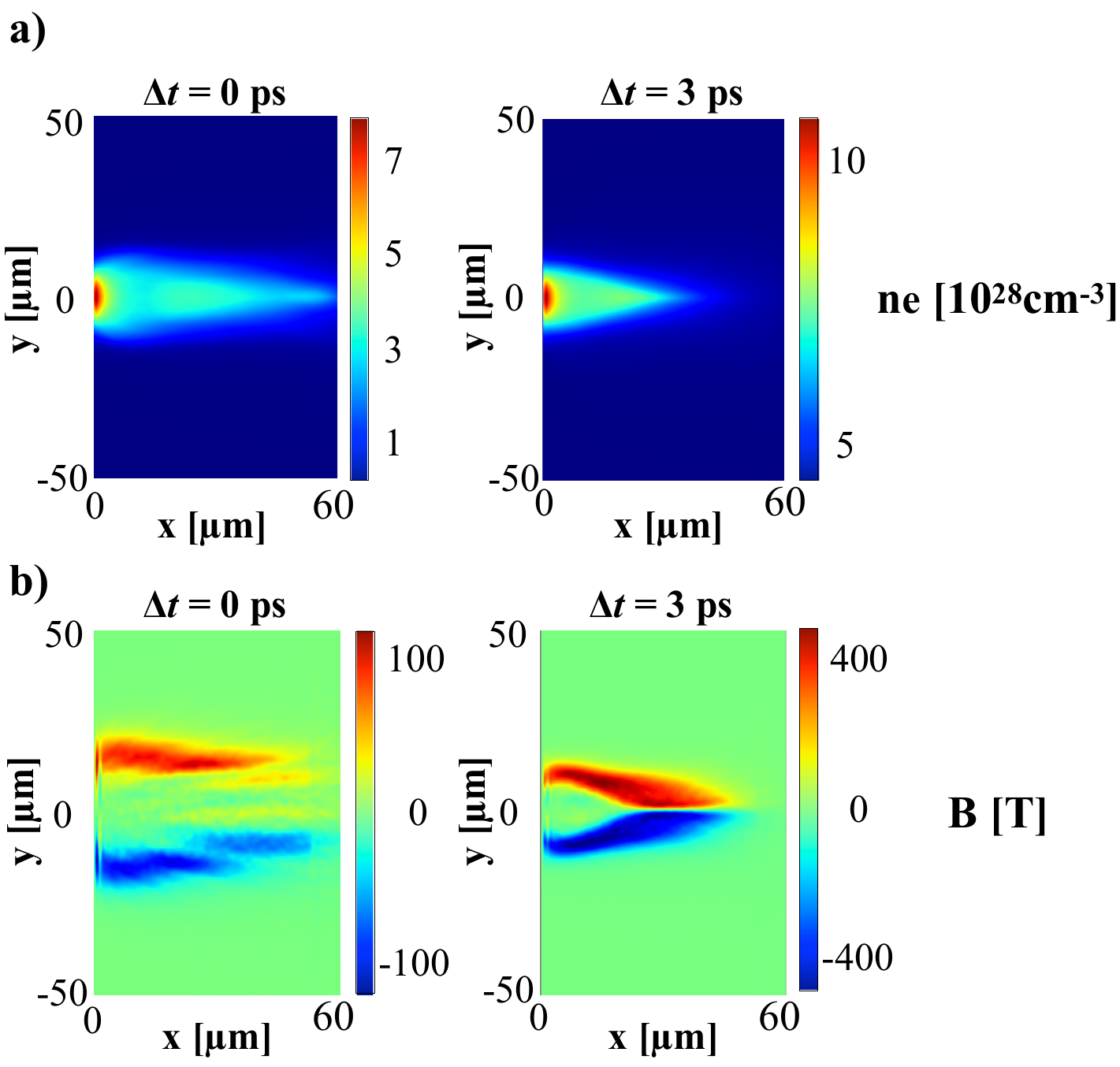}
\caption{\small (color online) Results of fast electron transport simulations for ${\varphi_{1}}/{\varphi_{2}}=2.5$. {\bf (a)} Comparison of the fast electron density profiles and {\bf (b)} of the azimuthal magnetic field component of magnetic field for $\Delta t = 0\,$ps (simultaneous interaction of the two laser pulses) and $\Delta t = 3\,$ps (optimum guiding of the fast electrons).}
\label{fig:Sim}
\end{figure}
\\
%
%
%
%
%
%
\\
\noindent In order to support our physical interpretation of the experimental results of fast electron collimation, we performed a set of numerical simulations. 
The pre-plasma formation by the interaction of the seed laser pulse with the front side aluminium layer was evaluated using the hydrodynamic code MULTI \cite{ramis1988} in 1D. The plasma electron density profile showed an approximately exponential profile which could be fitted as $n_e(x)\propto e^{ \frac{3}{2}x }$, where $x[\mu\textrm{m}]$ is the longitudinal coordinate. The parameters of the electron source produced by the interaction of the main laser pulse with a 50 $\mathrm{\mu}$m thick aluminium layer were evaluated via Particle-in-Cell (PIC) simulations in 2D using the SMILEI code \cite{smilei2000}. We considered a $470\,$ fs FWHM Gaussian pulse with $10^{19}\,$W.cm$^{-2}$ peak intensity. The extracted REB energy distribution  was averaged over the $1.5\,$ ps duration of the simulation and is well described in the $10\,$keV $\leq E \leq 200\,$MeV energy range by the following analytical expression: $f(E) = \exp{\left(-\frac{E} {T_b}\right)}+\left(\frac{T_c}{E}\right)\times \left[\frac {\left(\gamma_0 -1 \right) m_e c^2}{E}\right]^a \exp{\left(-\frac{E}{T_{sh}}\right)}$.
The fitting parameters are:  $T_b=30.3\,$keV, $T_{sh} = 10\,$MeV, $T_c = 1$, $\gamma_0 = 1.0075$, $a=1.6$.
The transport of fast electrons into the target was simulated in 3D with a hybrid-PIC code \cite{honrubia2009} using the aforementioned electron distribution as input to reproduces the experimental configuration for the run with a laser focal spot ratio ${\varphi_{1}}/{\varphi_{2}}=2.5$. The laser-to-fast-electrons conversion efficiency was set to $25\%$ according to \cite{solodov2008}. The electric resistivity is calculated using  the Eidmann-Chimier model \cite{chimier2007,eidmann2000}. The fast electron angular distribution is fitted by the function $f(\theta \,,x\,,E)\propto \exp\left[\left(-\frac{\theta -\theta_r}{\Delta \theta}\right)^2\right]$, where $\Delta \theta$ is the dispersion angle at the source and $\theta_{r}$ = actan[tan($\gamma$)$r/r_0$] is the mean radial angle with respect to the laser propagation axis with $r_0$= 13.5 $\mu$m. The angles for the first electron beam were chosen according to \cite{green2008} as  $\Delta \theta$ = $45^{0}$ and $\gamma$ = $35^{0}$, while the angles for the second electron source were determined from the PIC simulations as $55^{0}$ and $45^{0}$, respectively.
\\
The simulations for each delay time $\Delta t$ between laser pulses were run until the final time $t =3\,\textrm{ps}+\Delta t$. Results of hybrid simulations are compared with the experiment in terms of Cu-K$_{\alpha}$ emission spot sizes. As shown in figure \ref{fig:Scan2_image}(b) both the experimental and the synthetic Cu-K$_{\alpha}$ spot size exhibit a two times decrease at optimum delay when compared to a simultaneous shot of the two laser beams ($\Delta t = 0$). The minimum spot size is also reached with a delay time $\sim 3\,$ps, when the amplitude of magnetic field reached its maximum $B_{max}\sim400\,$T [Fig. \ref{fig:Sim}(b)]. The discrepancy in terms of size between the simulated and the experimental spots might be related to an underestimation of the fast electron beam divergence injected into hybrid simulations.
\\
\\
In summary, in the present experiment, we extensively studied the double pulse approach to the collimating of relativistic electron beams produced in high-intensity laser-plasma interactions. By changing the ratio between the focal spots of the two lasers ${\varphi_{1}}/{\varphi_{2}}$ and the injection time $\Delta t$, we observed a clear signature of collimation, validating the theory presented in \cite{robinson2008,volpe2014}. Two complementary diagnostic techniques have been implemented, which mainly show the respective behaviour of very fast vs. less fast hot electrons. Its experimental results are essentially in agreement. In particular, both from experimental results and from simulations, we have shown that for each value of ${\varphi_{1}}/{\varphi_{2}}$ there is an optimal injection time $\Delta t$, which, in agreement with expectations, increases when ${\varphi_{1}}/{\varphi_{2}}$ is decreased. We also found that about $70\%$ of hot electrons can be collimated by this mechanism. In conclusion, the double pulse technique appears to be an easy and controllable way to limit the divergence of fast electrons and improve energy transport deep into the target. This result opens interesting perspectives for a large variety of applications including the fast ignition approach to inertial confinement fusion and the optimisation of laser-driven particle sources.
\\

The research leading to these results has received funding from LASERLAB-EUROPE (grant agreement no. 654148, European UnionAos Horizon 2020 research and innovation programme). This experiment was carried out within the framework of the “Investments for the future” program IdEx Bordeaux LAPHIA (No. ANR-10-IDEX-03-02). Part of the used diagnostic equipment was funded by the French National Agency for Research (ANR) and the competitiveness cluster Alpha-Route des Lasers, project No. TERRE ANR-2011-BS04-014. Simulation work has been partially supported by the Spanish Ministry of Economy and Competitiveness (grant No. ENE2014-54960-R) and used HPC resources and technical assistance provided by Tirant and CeSViMa centers of the Spanish Supercomputing Network.The work was also supported by the Competitiveness Program of NRNU MEPhI, Russia. We gratefully acknowledge the support of the LULI ELFIE staff during the experiment campaign.

\bibliographystyle{unsrt}

\end{document}